\documentstyle[12pt]{article}
\begin{document}
\title{\bf Large Scale-Small Scale Duality and Cosmological Constant}
\author{{\bf Farhad Darabi}\\
{\small Department of Physics, Shahid Beheshti University, Evin 19839, Tehran, Iran.}\\
{\small Department of Physics, Tarbiyat Moallem University, Tabriz, P.O.Box 51745-406, Iran.}\\
{\small e-mail: f-darabi@cc.sbu.ac.ir}}
\maketitle
\begin{abstract}
We study a model of quantum cosmology originating from a classical model of
gravitation where a self interacting scalar field is coupled to gravity with
the metric undergoing a signature transition. We show that
there are dual classical signature changing solutions,
one at large scales and the other at small scales. It is possible to fine-tune
the physics in both scales with an infinitesimal effective cosmological
constant.
\end{abstract}

\vspace{15mm} 
\newpage

\section*{1. Introduction}

The question of signature transition in classical and quantum gravity has
been the subject of intense investigations during the past few years. These
investigations have led to, basically, two approaches to the subject, discrete
and continuous. The continuous approach has been studied by a number of
authors. Of particular interest to the present work is the model adopted by
Dereli and Tucker \cite{DT} in which a self interacting scalar field is
coupled to gravity. In this model, Einstein's field equations, coupled to
the scalar field, are solved such that the scalar field and the scale factor
are considered as dynamical variables, giving rise to cosmological solutions
with degenerate metrics, describing transition from Euclidean to Lorentzian
domains. A quantum cosmological version of this model was used to derive the
wavefunction of the universe by solving the corresponding Wheeler-DeWitt
equation \cite{DOT}. This was achieved by adopting a new choice of
variables through which Einstein's classical equations of motion arise from
an anisotropic constrained oscillator-ghost-oscillator Hamiltonian. A family
of Hilbert subspaces were then derived in which states are identified with
non dispersive wave packets peaking in the vicinity of classical loci with
parametrizations corresponding to metric solutions of Einstein's equations
that admit a continuous signature transition.

In the approach detailed below we use the notations in \cite{DT} and \cite{DOT}
with their results and pay
attention to duality transformations on the couplings in the present scalar
field potential such that the classical signature changing cosmology transforms to its
``dual'', but the quantum cosmology remains unchanged. As a result of \cite{DOT}
, a remarkable correlation should be seen between this self dual quantum
cosmology and dual classical cosmologies. Based on this correlation, we can introduce
the ``large scale-small scale'' duality interrelated with the ``small coupling-large coupling''
duality in a way that dual classical cosmologies, one at large scales and the other at small
scales, may have exactly a same infinitesimal effective cosmological constant.
It was suggested \cite{all} that a large distance-small distance connection should exist
to shift the cosmological constant to zero and wormholes provide such a connection.
Therefore, such a fine-tuning in the present model will be of particular interest since
it may explain that:
{\em why the cosmological constant should be small from the microscopic point of view
if it is small from the cosmic viewpoint} \cite{all}. Moreover the fine-tuning mechanism is
related, somehow, to the existence of quantum wormholes in the model.

\section*{2. Dual classical cosmologies}

Consider the Einstein-Hilbert action
\begin{equation}
I=\int_{}^{}\!\sqrt{|g|}\: \left[\frac{1}{16 \pi G} {\cal R} + {\cal L}_
{\em matter}\right] d^{4}x
\end{equation}
where ${\cal L}_{\em matter}=\frac{1}{2} \partial_{0}\phi\:
\partial^{0}\phi - U(\phi)$ is the real scalar field Lagrangian (assume that
$\phi$ is homogeneous and only depends on time). We choose the chart $
\{\beta,x^1,x^2,x^3\}$ and parametrize the metric as \cite{DT}
\begin{eqnarray}
g = -\beta d\beta \otimes d\beta + \frac{R^2 (\beta)}{[1+{\frac{k}{4} r^2}]^2} \sum_i d{x^i} \otimes d{x^i} \label{eq3}
\end{eqnarray}
where $r^2=\sum x_ix^i$, $R(\beta )$ is the scale factor with $k=\{-1,0,1\} $
representing open, flat or closed universes and $\beta $ is the lapse
function with the hypersurface of signature change at $\beta =0$. For $\beta
>0$, the cosmic time can be written as $t=\frac 23\beta ^{3/2}$. The scalar
curvature ${\cal R}$ is then given by
\begin{equation}
{\cal R}= 6 \left[\frac{\ddot{R}}{R} + \frac{k+{\dot{R}}^2}{R^2}\right]
\end{equation}
where $.$ denotes the derivetives with respect to $t$ and the units is taken such that $3\pi G=1$. Making the transformations \cite
{DT}
\begin{eqnarray}
X = R^{3/2} \cosh(\alpha \phi)  \label{eq8}\\
Y = R^{3/2} \sinh(\alpha \phi)  \label{eq9}
\end{eqnarray}
where $-\infty <\phi <+\infty ,\,\,\,0\leq R<\infty $ and $\alpha ^2=\frac
{3}{8} $, the corresponding effective Lagrangian form is obtained
\begin{eqnarray}
2{\alpha}^2 {\cal L} dt = \left\{-\dot{X}^2 + \dot{Y}^2 + \frac{9k}{4} (X^2 -Y^2)^{1/3} - 2 \alpha^2 (X^2 - Y^2)
U(\phi(X,Y))\right\} dt. \label{eq10}
\end{eqnarray}
Applying the Einstein equations,
one can see that the Hamiltonian ${\cal H}$ corresponding to ${\cal L}$,
must vanish identically giving a ``zero energy condition'' to be imposed on
the solutions of the field equations. Concentrating on cosmologies with $k=0$%
, we take the potential as
\begin{eqnarray}
2 \alpha^2 (X^2 - Y^2) U(\phi(X,Y)) = a_1 X^2 + (a_2 - a_1^{-1}) Y^2 + 2bXY  \label{eq11}
\end{eqnarray}
where $a_1,a_2$ and $b$ are assumed to be dimensionless constant parameters
\footnote{%
Such an assumption is reasonable in the plank units}. As we shall see, this
particular choice of the potential \footnote{%
This potential is the same as that of Ref. 1 except a little difference in
the coefficient of $Y^2$.} leads to what we shall call ``duality
transformations''. Variation of the action with respect to the dynamical
variables X and Y now gives
\begin{eqnarray}
\ddot{X} &=& a_1 X+b Y  \label{eq12}\\
\ddot{Y} &=& -b X + (a_1^{-1}-a_2)Y  \label{eq13}
\end{eqnarray}
for which the ``zero energy'' condition are to be imposed. Writing the
potential in terms of the physical couplings $\lambda $ , $m^2$ and $b$ we obtain
\cite{DT}
\begin{equation}
\label{eq14}U(\phi )=\lambda +\frac 1{2{\alpha }^2}m^2\sinh ^2(\alpha \phi
)+\frac 1{2{\alpha }^2}b\sinh (2\alpha \phi )
\end{equation}
where $\lambda =U\mid _{\phi =0}=\frac{a_1}{2\alpha ^2}$, $m^2=\frac{{%
\partial }^2U}{\partial {\phi }^2}\mid _{\phi =0}=a_1+a_2-a_1^{-1}$ and $b$
are the bare cosmological constant, positive mass square and coupling
constant respectively. The minimum of the potential occurs when $|{2b/m^2}%
|<1 $. Equations (\ref{eq12}) and (\ref{eq13}) may be decoupled into normal
modes $\alpha=(
\begin{array}{c}
u \\
v
\end{array}
)$ with ``zero energy'' solutions for $\lambda _{+},\lambda _{-}<0$ given by
\cite{DOT}
\begin{eqnarray}
v&=& 2 A_0 \cos\left[ \frac{1}{r} \mbox{cos}^{-1}\left(\epsilon\frac{ru}{2A_0}\right)\right], \hspace{4mm}
|u|\leq\frac{2A_0}{r} \nonumber\\
   \label{eq15}\\
v&=& 2 A_0 \cosh\left[ \frac{1}{r} \mbox{cosh}^{-1}\left(\epsilon\frac{ru}{2A_0}\right)\right], \hspace{4mm} |u|>\frac{2A_0}{r}
\nonumber
\end{eqnarray}
where $\lambda _{\pm }$ are the eigenvalues of the matrix
$(\begin{array}{cc} a_1&b \\-b&a_1^{-1}-a_2 \end{array})$ given by
\begin{eqnarray}
\lambda_{\pm}=\frac{a_1+a_1^{-1}-a_2}{2}\pm\sqrt{\left(\frac{a_1+a_1^{-1}-a_2}{2}\right)^2-(1-a_1a_2+b^2)}  \label{eq16}
\end{eqnarray}
with $r=\sqrt{\frac{\lambda _{+}}{\lambda _{-}}}$ , $0<r<1$, $A_0$ an
arbitrary real constant and $\epsilon =\pm 1$ indicating two ways to satisfy
the Hamiltonian constraint ${\cal H}=0$.

An interesting feature of this model is that one can find a class of
transformations on the space of parameters $\{a_1,a_2,b\}$ which would leave
the eigenvalues $\lambda _{\pm }$ invariant. These
transformations can be written as
\begin{eqnarray}
& &a_1 \rightarrow a_1^{-1} \nonumber\\
& &a_2 \rightarrow a_2  \label{eq17}\\
& &b^2 \rightarrow b^2 + (a_1^{-1} - a_1) a_2 . \nonumber
\end{eqnarray}
In terms of the physical couplings $\lambda $, $m$ and $b$, the eigenvalues
are
\begin{equation}
\label{eq18}\lambda _{\pm }=\frac{3\lambda }4-\frac{m^2}2\pm \frac 12\sqrt{%
m^4-4b^2} .
\end{equation}
It is seen that although the classical loci (\ref{eq15}) on the configuration
space $(u,v)$ remains unchanged under
(\ref{eq17}), the corresponding solutions $R(\beta )$ and $\phi (\beta )$
change, since $X(\beta )$ and $Y(\beta )$ are related to $u(\beta )$ and $%
v(\beta )$ by the decoupling matrix which changes under (\ref{eq17}) (see (\ref{b}, \ref{c}).
In terms of the physical couplings, (\ref{eq17}) can be rewritten as
\begin{eqnarray}
& &\lambda \rightarrow \tilde{\lambda} \equiv \frac{1}{4 \alpha^4} \lambda^{-1} \nonumber\\
& &m^2 \rightarrow \tilde{m}^2 \equiv m^2 - \frac{4 \alpha^4 \lambda^2 -1}{\alpha^2 \lambda} \label{eq18}\\
& &b^2 \rightarrow \tilde{b}^2 \equiv b^2 + m^2 [(2 \alpha^2 \lambda)^{-1} - 2 \alpha^2 \lambda] + [(2 \alpha^2 \lambda)^{-1} -
2 \alpha^2 \lambda]^2 .\nonumber
\label{O}
\end{eqnarray}
Therefore, if we define (\ref{eq18}) as ``duality'' transformations, then we
have two sets of solutions for $R(\beta )$ and $\phi (\beta )$ corresponding
to dual sets of physical couplings. We interpret the new couplings as dual
bare cosmological constant, dual mass square and dual coupling constant
respectively.

The behavior of the physical couplings under the duality transformations
merits some discussion. As is discussed in \cite{DT}, in order to have
signature transition, both eigenvalues $\lambda _{\pm }$ must be negative
and equation (14) gives
\begin{equation}
\lambda <\frac 43[\frac{m^2}2-\frac 12\sqrt{m^4-4b^2}] .
\label{11}
\end{equation}
However, this does not guarantee that the dual potential has also a minimum and a
positive mass square. In order to have both we take
\begin{equation}
\label{eq19}\frac{m^2}2-\frac 12\sqrt{m^4-4b^2}\leq 1 .
\label{22}
\end{equation}
One can find a suitable choice of couplings satisfying (\ref{eq19}) so as to make the
dual potentials $U(\phi)$ and ${\tilde U}(\tilde\phi)$ have the same properties in a way that once
$\mid2 b/{m}^2\mid <1$, then $\mid2\tilde b/{\tilde m}^2\mid <1$, ensuring that
both dual potentials have minimum (see below).
In the case of a small cosmological constant the dual transformations map the
desired small values of the bare cosmological constant $\lambda $, positive mass
square $m^2$ and coupling constant $b$ to large values of the corresponding
dual couplings, $\tilde\lambda $, ${\tilde m}^2$ and $\tilde b$.
It then follows that two different classical cosmologies, one with
small couplings $\{\lambda, m^2, b\}$ and the other with large dual couplings
$\{\tilde\lambda,{\tilde m}^2, \tilde b\}$ exhibit the same signature dynamics
on the configuration space $(u,v)$.

It is straightforward to show that the effect of ``small $\rightarrow$ large''
couplings transformations on the space of physical solutions $R(\beta)$ and
$\phi(\beta)$ may cause a ``large $\rightarrow$ small'' scales transformations
for a suitably chosen set of couplings.
To see, one may recover $R$ from $X$ and $Y$ as
\begin{equation}
R = (X^2 - Y^2)^{\frac{1}{3}}
\label{a}
\end{equation}
on the other hand, we have the following change of variables
\begin{equation}
\xi={\cal S} \alpha
\label{b}
\end{equation}
where $\xi=(
\begin{array}{c}
X \\
Y
\end{array}
)$ and
\begin{equation}
{\cal S}=\left(\begin{array}{cc} -\frac{m^2-\sqrt{m^4-4b^2}}{2b} & -\frac{m^2+\sqrt{m^4-4b^2}}{2b}\\ \\1 & 1 \end{array}\right)_.
\label{c}
\end{equation}
Now, by choosing the set of small couplings as
\footnote{Here $m$ as a reasonably {\em small} mass may be the planck mass $M_{pl}$
, or another {\em small} energy density related to some spontaneous symmetry-breaking
scale such as $M_{SUSY}$ or $M_{weak}$. On the contrary, consistent with equations
(\ref{11}) and (\ref{22}), the two other couplings $\lambda$ and $b$ are assumed to
be very small compared to $m^2$ namely, $m^2\gg b, \lambda$.}
\begin{equation}
\lambda \simeq 0 \hspace{20mm}m^2 \ll 1 \hspace{20mm} b \simeq 0
\label{d}
\end{equation}
with the assumption that $m^2 \gg b, \lambda$ we find from (\ref{a}-\ref{c}) that $R$ may have a desired large value due to the
adjustable large value of the matrix element $|{\cal S}_{12}|\simeq \frac{m^2}{b} \gg 1$.
On the other hand, taking the dual set of couplings through (15) we find the set of large
couplings $\{\tilde{\lambda}, \tilde{m}^2, \tilde{b}\}$ for which the corresponding
decoupling matrix (\ref{c}) leads through (\ref{a},\ref{b}) to an small value for $R$.
This is because $\tilde{m}^4-4\tilde{b}^2=m^4-4b^2 \ll 1$ and
$\tilde{m}^2 \sim \tilde{b} \sim \lambda^{-1} \gg 1$,
so $|\tilde{{\cal S}}_{11}| \sim |\tilde{{\cal S}}_{12}| \sim 1$.

This is a remarkable step towards solving the cosmological constant problem in the
context of the present model.
The cosmological constant plays two roles in physics. The first one, as a coupling
constant in microscopic physics, has its origin in short distance physics whereas
the other role, as a macroscopic parameter, controls the large scale behavior
of the universe. There is no explanation of why the cosmological constant is so small
that the universe can be flat and big enough.
A direct connection between the large scale physics and the small scale physics
should exist by which the microscopic physics be fine-tuned with good precision
so that the large scale structure of space-time can look as it is observed today.
One possible connection is due to the wormholes which provide such a large scale -
small scale relation.

Peoples \cite{all} have considered the effects of wormholes in the Euclidean path integral of
quantum gravity and shown that the effect is modifying the cosmological constant
and providing a probability distribution concentrating at zero value of the cosmological constant.
In particular, Hawking proposed that quantum fluctuations in space-time topology
at small scales may play an important role in shifting the cosmological constant to zero.

The appearance of ``large scale-small scale'' duality interrelated with ``small coupling-large
coupling'' duality in the present model is a good chance to
obtain a fine-tuning mechanism to have an infinitesimal effective cosmological constant at
both large and small scales. For the sake of definiteness we define the minima
of the potentials $U(\phi)$ and ${\tilde U}(\tilde\phi)$ respectively \cite{DT}
\begin{equation}
\Lambda_{eff} = \lambda+\frac{m^2}{4\alpha^2}\left(\sqrt{1-\frac{4 b^2}{m^4}}-1\right),
\label{44}
\end{equation}
\begin{equation}
{\tilde\Lambda}_{eff} = \tilde\lambda+\frac{{\tilde m}^2}{4\alpha^2}\left(\sqrt{1-\frac{4 {\tilde b}^2}{{\tilde m}^4}}-1\right)
\label{55}
\end{equation}
as dual effective cosmological constants in the form of the sum of two terms, the
bare cosmological constant and the contribution of mass scale of the scalar field.
Thanks to the duality transformations (15) we find the remarkable result
\begin{equation}
\Lambda_{eff}=\tilde{\Lambda}_{eff}
\label{33}
\end{equation}
showing the self-duality of the effective cosmological constant.
We now choose the set of suitably small couplings
$\{\lambda, m^2, b\}$ corresponding to the large scale $R(\beta)$
\begin{equation}
\lambda \simeq 0 \hspace{20mm}m^2 \ll 1 \hspace{20mm} b \simeq 0
\end{equation}
with $b \ll m^2$. Therefore, according to (\ref{44}), the effective cosmological
constant at large scales becomes infinitesimal, namely
\begin{equation}
\Lambda_{eff} \simeq 0 .
\end{equation}
At small scales ${\tilde R}(\beta)$, however, due to duality transformations (15)
the dual couplings
$\{\tilde\lambda,{\tilde m}^2, \tilde b \}$ are then very large and we have really a
significant potential ${\tilde U}(\tilde \phi)$ with strong couplings which are
anticipated to affect essentially the value of the cosmological constant giving rise to the
well known cosmological constant problem. However, as is shown in (\ref{33}), the
duality transformations are such that at small scale we have exactly the same
infinitesimal effective cosmological constant
\begin{equation}
\tilde{\Lambda}_{eff}=\Lambda_{eff} \simeq 0 .
\end{equation}
Therefore, both $\Lambda_{eff}$ at large scales and ${\tilde\Lambda}_{eff}$ at
small scales can be fine-tuned to zero by the same procedure including
duality transformations. In other words, these duality transformations provide a
direct connection between the large and small scales and set the cosmological
constant, at both scales, to zero.

In the next section, we will see that the origin of these dual signature changing
classical cosmologies with a {\em self-dual} effective cosmological constant
is in a {\em self dual} Wheeler-DeWitt equation with dual solutions
obeying the wormhole boundary conditions.

\section*{3. Self Dual Quantum Cosmology- Dual Quantum Wormholes}

In the previous section we have shown that it is possible to find dual
classical cosmologies on the ($R,\phi $) configuration space corresponding
to a unique classical cosmology defined on the ($u,v$) configuration space.
On the other hand it is shown \cite{DOT} that the corresponding
Wheeler-Dewitt equation in terms of variables ($u,v$) is
\begin{equation}
\left\{\frac{\partial ^2}{\partial u^2}-\frac{\partial ^2}{\partial v^2}-\omega
_1^2u^2+\omega _2^2v^2\right\}\Psi (u,v)=0
\end{equation}
where $\omega _1^2=-\lambda _{+},\omega _2^2=-\lambda _{-}$. It has
oscillator-ghost-oscillator solutions belonging to the Hilbert space ${\cal H%
}^{(m_1,m_2)}({\cal L}^2)$ as
\begin{equation}
\Psi ^{(m_1,m_2)}(u,v)=\sum_{l=o}^\infty \!c_l\Phi _l^{(m_1,m_2)}(u,v)
\label{psi}
\end{equation}
with $m_1,m_2\ge 0$ and $c_l\in C$. The basis solutions $\Phi
_l^{(m_1,m_2)}(u,v)$ are separable as
\begin{equation}
\Phi _l^{(m_1,m_2)}(u,v)=\alpha _{m_2+(2m_2+1)l}(u)\beta _{m_1+(2m_1+1)l}(v)
\end{equation}
with normalized solutions
\begin{equation}
\begin{array}{l}
\alpha _n(u)=(
\frac{\omega _1}\pi )^{1/4}\frac{\exp {-\frac{\omega _1u^2}2}}{\sqrt{2^nn!}}%
H_n(\sqrt{\omega _1}u) \\  \\
\beta _n(v)=(\frac{\omega _2}\pi )^{1/4}\frac{\exp {-\frac{\omega _2v^2}2}}{%
\sqrt{2^nn!}}H_n(\sqrt{\omega _2}v)
\end{array}
\end{equation}
where $H_n(x)$ are {\em Hermite} polynomials. Obtaining these solutions requires a
quantization condition \cite{DOT}
\begin{equation}
\frac{\lambda _{+}}{\lambda _{-}}=\left(\frac{2m_1+1}{2m_2+1}\right)^2
\end{equation}
which is imposed on the couplings in the scalar field potential. For a
given pair of ($m_1,m_2$) it is shown \cite{DOT} by graphical analysis that
the absolute value of the solutions (\ref{psi}) have maxima in the vicinity of
classical loci (11) admitting a signature transition. This defines a definite
correlation between classical loci and quantum solutions on the space of the
$(u,v)$ variables. On the other hand, we have already shown that
there are duality transformations on the couplings in a given scalar
field potential giving rise to a dual potential such that the solutions of
field equations on the ($R,\phi $) configuration space transform to dual
solutions $(\tilde R,\tilde \phi)$ whereas the solutions on the ($u,v$) configuration
space remain unchanged. Applying the quantum cosmology discussed above to this picture,
it turns out that for any pair of ($m_1,m_2$) defining a distinct quantum cosmology
in terms of the variables ($u,v$), we may correspond dual classical
solutions ($R,\phi $) and ($\tilde R,\tilde \phi $) admitting signature
transition from a Euclidean to a Lorentzian space-time. This is because starting separately
with dual classical systems $U(\phi)$ and ${\tilde U}(\tilde \phi)$ tends to the
same quantization condition and the same quantum cosmology concentrated on both over the configuration
space ($u,v$).

In this respect, for a given pair $(m_1,m_2)$ we have a quantum cosmology with dual predictions: a large scale
cosmology $R(\beta)$ and an small scale cosmology ${\tilde R}(\beta)$, both
having an infinitesimal effective cosmological constant. Time independence
of quantum cosmology \footnote{Wheeler-DeWitt equation as zero energy schrodinger
equation gives rise to time independent solutions.} would let us to choose two different
parametrization of classical loci (11) in terms of classical time \cite{DOT}
so that the large scale cosmology is an old signature changing universe and
the small scale one being the microscopic fluctuation in space-time signature with
a short life-time.

According to a common belief \cite{all}, {\em the microscopic fluctuations in space-time
topology or the microscopic wormholes may play an important role in shifting the cosmological constant to zero.}
This idea may be realized, somehow, in the present model for two reasons: firstly,
the phenomenon of dynamical space-time topology change may be accompanied by a
dynamical signature change of space-time metric and in the present model ( see (\ref{55}) )
we see that at small scale ${\tilde R}(\beta)$, the large bare cosmological
constant $\tilde\lambda$ can only shift to zero if there exist a signature changing
mechanism \cite{DT}; secondly, there are quantum wormholes corresponding
to the matter field potentials $U(\phi)$ and its dual ${\tilde U}(\tilde\phi)$
\footnote{A classical scalar field potential as
$U(\phi)=\frac{2}{9}m^2 \sinh^2 \frac{3}{2} \phi$ is also introduced by Hawking
and Page \cite{HP} giving rise to quantum wormholes. That potential is the special
form of the present one $U(\phi)$ with $b=0$.}.
In both cases, it is easy to show that the resulting quantized model of signature transition may give
exponential wave functions ( one with large variable $R$ and the other with small
variable $\tilde R$ ) satisfying the Hawking-Page boundary conditions for
the existence of quantum wormholes
\footnote{The conditions are: {\em Wave function should decay exponentially for large scale factor
and that be well behaved when the scale factor goes to zero.}
An attempt to obtain the quantum wormholes in the context of classical signature
change \cite{ccs} shows that for any choice of perfect fluid matter, there are no quantum
wormhole solutions while it may be there some classical wormholes.} \cite{HP}.
Indeed, because the Wheeler-DeWitt equation is independent of the lapse function
$\beta$, the Euclidean regime as well as the Lorentzian one is already included
in the formalism. So the exponential solutions are to be expected as well as
oscillating solutions.
One should then interpret the
large scale quantum wormholes as macroscopic and the small scale ones as microscopic.
Therefore, the existence of microscopic quantum wormholes in this model in which
the cosmological constant can be tuned to zero at small scale, may be a realization of
the common belief mentioned above.

\newpage
{\Large {\bf Conclusion} \vskip8pt\noindent}

We have shown that in the context of signature transition in
classical and quantum cosmology based on the models proposed by Dereli
{\em et al.} one may find, by appropriate choice of the couplings in the scalar
field potential, a set of duality transformations such
that relate the dual classical solutions to each other, while a same quantum
cosmology is concentrated equally on both of them. In this respect we may have a
quantum cosmology correlated with dual classical predictions. The nice property
of duality transformations, based on this correlation between quantum and classical
cosmologies, makes an interrelation between ``large scale-small scale'' duality and
``small coupling-large coupling'' duality. Thanks to this interrelation we have shown
that it is possible to fine-tune the physics at both large and small scales to
have exactly a same infinitesimal effective cosmological constant.
In other words: {\em A choice of an infinitesimal cosmological constant at microscopic scales
implies an infinitesimal cosmological constant at cosmic scales}.\\
The interesting features of this model are:

1) There is no semi-classical approximation in relating the solutions of Wheeler-DeWitt
equation to dual classical solutions \cite{DOT}, hence we are not worried about the
breakdown of any semi-classical approximation at small scales except planck scale
at which neither Einstein nor Wheeler-DeWit equations make sense. On the other hand,
there is an alternative possibility: {\em if one assumes this Large scale-small scale
duality potentially fundamental, then it should connect the typical size of the
universe to a dual ``natural'' size  which is not necessarily of planck size.}
This will let us to avoid the need for a quantum gravity domain. Recently, such
ideas has been the subject of intense investigations \cite{Nima}.

2) Potentially, there is an evidence of topology change through the existence of
quantum wormholes in the context of classical signature change.

3) It is seen that the wave packets, constructed by the quantum wormhole solutions
of the Wheeler-DeWitt equation peak in the vicinity of dual classical cosmologies
with a nice property: {\em having a self-dual fine-tunable effective cosmological constant}.
The fact that a {\em self-dual} quantum cosmology peaks about a {\em self-dual}
fine-tunable effective cosmological constant is of particular importance since
it explains that the resolution of the cosmological constant problem may be in
the quantum cosmological considerations.

4) The fine-tuning of effective cosmological constant at large and small scales
could lead to a deeper understanding concerning the origin of the arrow of time
at large and small scales, since the fine-tuning mechanism here is based on the
signature transition ``from'' Euclidean (no time) ``to'' Lorentzian (beginning of time) domains.

5) From equation (\ref{33}) we find that for a chosen set of couplings
\{$\lambda, m^2, b$\} and its dual set \{$\tilde{\lambda}, \tilde{m}^2, \tilde{b}$\}
a {\em self-dual} effective cosmological constant is attributed and these dual sets
may correspond to dual typical scales. Therefore, one may fine-tune the corresponding effective
cosmological constant for any dual scales emerging from dual sets
\{$\lambda, m^2, b$\} and \{$\tilde{\lambda}, \tilde{m}^2, \tilde{b}$\}.
\newpage
{\Large {\bf Acknowledgment} \vskip8pt\noindent}

I would like to thank H. R. Sepangi and H. Salehi for useful discussions.


\begin{thebibliography}{99}
\bibitem{DT} T. Dereli, R. W. Tucker, Class. Quantum Grav. {\bf 10},
365 (1993).
\bibitem{DOT} T. Dereli, M. \"{O}nder, R. W. Tucker, Class.
Quantum Grav. {\bf 10}, 1425 (1993).
\bibitem{HP} S. W. Hawking, D. N. Page, Phys. Rev. {\bf D 42}, 2655
(1990).
\bibitem{all}S. Hawking, Nucl. Physics. {\bf B 144}, 349 (1978);
Phys. Lett. {\bf B 134}, 403 (1984); \\Phys. Lett {\bf B 195}, 337 (1987).\\
A. Linde, Phys. Lett. {\bf B 200}, 272 (1988);\\
I. Klebanov, L. Susskind and T. Banks, Nucl. Physics. {\bf B 317}, 665 (1989);\\
G. Horowitz, M. Perry and A. Strominger, Nucl. Physics. {\bf B 238}, 653 (1984);\\
A. Strominger, Phys. Rev. Lett. {\bf 52}, 1733 (1984);\\
S. Giddings and A. Strominger, Nucl. Physics. {\bf B 306}, 890 (1988); {\bf B 307}, 584 (1988); \\
{\bf B 321}, 481 (1988)\\
S. Coleman, Nucl. Physics. {\bf B 310}, 643 (1988).
\bibitem{ccs}A. Carlini, D. H. Coule, D. M. Solomons, Mod. Phys. Lett. {\bf A 11}, 1453 (1996).
\bibitem{Nima}Nima Arkani-Hamed, Savas Dimopoulos, Nemanja Kaloper, John March-Russell,\\ hep-ph/9903239;
hep-ph/9903224.\\
Nima Arkani-Hamed, Savas Dimopoulos, Gia Dvali, Phys. Rev. {\bf D 59}, 086004 (1999); Phys. Lett. {\bf B 429}, 263 (1998).
\end{thebibliography}
\end{document}